\documentclass[runningheads,empty]{jmlr}

\usepackage{natbib}
\usepackage{lipsum}
\usepackage{graphicx}
\usepackage{amsmath}
\usepackage{float}
\usepackage{svg}
\setcitestyle{numbers}
\usepackage{booktabs}
\usepackage{longtable}
\usepackage{siunitx}
\usepackage{fancyhdr} 

\def\set@curr@file#1{\def\@curr@file{#1}}
\makeatother

\jmlrvolume{[VOLUME \# TBD]}
\jmlryear{2024}

\begin{document}

\title[AFEN: Respiratory Disease Classification using Ensemble Learning]{AFEN: Respiratory Disease Classification using Ensemble Learning}

\author{\Name{Rahul Nadkarni}
       \Email{rnadkarn@ucsc.edu}\\ 
       \addr Computer Science and Engineering Department\\
       University of California, Santa Cruz\\
       Santa Cruz, CA, United States of America
       \AND
       \Name{Emmanouil Nikolakakis}
       \Email{enikolak@ucsc.edu}\\ 
       \addr Electrical and Computer Engineering Department\\
       University of California, Santa Cruz\\
       Santa Cruz, CA, United States of America
       \AND
       \Name{Razvan Marinescu}
       \Email{ramarine@ucsc.edu}\\ 
       \addr Computer Science and Engineering Department\\
       University of California, Santa Cruz\\
       Santa Cruz, CA, United States of America
       } 

\rhead{Rahul Nadkarni et al.}
\maketitle

\begin{abstract}
\hspace{1em} We present AFEN (Audio Feature Ensemble Learning), a model that leverages Convolutional Neural Networks (CNN) and XGBoost in an ensemble learning fashion to perform state-of-the-art audio classification for a range of respiratory diseases. We use a meticulously selected mix of audio features which provide the salient attributes of the data and allow for accurate classification. The extracted features are then used as an input to two separate model classifiers 1) a multi-feature CNN classifier and 2) an XGBoost Classifier. The outputs of the two models are then fused with the use of soft voting. Thus, by exploiting ensemble learning, we achieve increased robustness and accuracy. We evaluate the performance of the model on a database of 920 respiratory sounds, which undergoes data augmentation techniques to increase the diversity of the data and generalizability of the model. We empirically verify that AFEN sets a new state-of-the-art using Precision and Recall as metrics, while decreasing training time by 60\%.
\end{abstract}

\section{Introduction}

\hspace{1em}

Respiratory diseases affect millions of people globally every year, claiming 4 million lives annually\cite{chen_global_2023} and diagnosing around 450 million individuals at any given point. While these diseases target the same human body system, their severity can vary from mild to life-threatening. Therefore, accurate diagnosis methods must be available to ensure patients receive the correct treatment.

Today's methodologies for diagnosing respiratory diseases heavily rely on accurately mapping lung auscultations to the correct diseases. Common diagnostic processes can be invasive and can introduce harmful radiation into human body systems \cite{jung_early_2021}. To address this issue, audio classification can enable efficient, non-invasive diagnosis of respiratory diseases. Machine learning models have the capability to analyze waveforms and detect patterns beyond the human ear's perception. Additionally, they can provide medical practitioners with the means to validate their diagnoses or justify the use of radiation.

Convolutional Neural Networks (CNNs) have been a staple in the field for analyzing Mel-Spectrograms, and much of the architecture has remained consistent for years. Additionally, common feature extraction methods are limited to Mel Frequency Cepstral Coefficients (MFCCs) and Mel Spectrograms. While they are able to execute a holistic signal analysis, current works do not take advantage of the full spectrum of features at their disposal. This limited complexity has resulted in higher training times coupled with lower accuracy.

In this work, we introduce AFEN (Audio Feature Ensemble Learning) a novel Ensemble Learning method for Respiratory Disease classification. We utilize a Multi-Feature CNN that takes audio features as input and concatenates their result. Secondly, we utilize the XGBoost \cite{xgboost} model on respective features before fusing both models using soft voting. The results set a new state of the art for this architecture and for this dataset in accuracy.

\subsection*{Generalizable Insights about Machine Learning in the Context of Healthcare}

Our main contributions are as follows:

\begin{enumerate}
    
    \item \textbf{Feature Extraction}: We devised a unique feature extraction strategy that captured relevant information from the augmented audio data. We targeted a new and expanded set of features, including Mel-Frequency Cepstral Coefficients (MFCC), Mel Spectrograms, Chroma Short Time Fourier Transforms (CSTFT), Spectral Rolloff, and the Zero Crossing Rate.
    
    \item \textbf{Model Training with XGBoost and CNNs}: We trained two separate models using the extracted features. We chose XGBoost for its interpretability and its ability to handle features effectively, while we selected CNNs for their capability to learn features directly from spectrograms or time-frequency representations of the audio signal. Finally, we used ensemble learning techniques to fuse the predictions of both models, achieving a more comprehensive classification performance than that of the individual models.
    \item \textbf{Performance Evaluation and Comparison}: We evaluated the performance of the proposed ensemble model against existing methods on this dataset. We utilized appropriate performance metrics for classification, such as precision, recall, and AUC, for a comprehensive evaluation of the model. Our focus was on the significant increase in accuracy across all classes. Additionally, we showcased the practical feasibility and scalability of the proposed method for real-world deployment by reducing the number of epochs compared to current works.
\end{enumerate}
\newpage
\section{Related Work}

\hspace{1em} 

While Natural Language Processing and Computer Vision dominate the current technological landscape, Audio Machine Learning Applications remain relatively unexplored. Within the Audio Machine Learning field, there exists two main disciplines: Classification and Generation. Classification is the most common application of Audio Machine Learning, with Generation still growing. Studies such as \cite{gunduz_deep_2019}, \cite{scarpiniti_deep_2021}, and \cite{lu_content-based_2003} provide valuable approaches into classification techniques. Similarly, studies like \cite{kong_hifi-gan_2020}, \cite{mehri_samplernn_2016}, and \cite{kreuk_audiogen_2023} have all contributed to the field of audio generation.

Given the correlation between sound and Respiratory Diseases, it was a natural progression to utilize Machine Learning techniques to classify diseases. In 2017, \cite{pramono_automatic_2017} conducted a survey of the current audio machine learning landscape and found that no clear standard benchmark had been set. They concluded that Support Vector Machines would be a suitable approach to respiratory disease detection for its versatility to be used for binary or multi-class classification. However, \cite{aykanat_classification_2017} compared the accuracy of a Convolutional Neural Network to Support Vector Machines. They tested both on binary classification between healthy and diseased sounds, and multi-class classification for auscultation sounds. Using 2 convolutional layers, they achieved results similar to their Support Vector Machine. This work was one of many to shift the baseline from SVMs to CNNs by maintaining a high level of performance while reducing the model's complexity.

In keeping with this standard, \cite{song_ahi_2023} introduced a new architecture for Ensemble Learning in Audio Classification. They extracted MFCCs, Perceptual Linear Predictives (PLPs), Subband Energy, Gammatone Frequency Cepstral Coefficients (GFCCs), Crest Factor, and a variety of Spectral features to train their Ensemble Network. Their network consists of a CNN and a ResNet-18 to train Mel Spectrum features and an XGBoost Model for each acoustic feature. The models are trained separately but culminate in their fusion to generate an Apnea-HypoApnea Index score for patients facing Obstructive Sleep Apnea-Hypoapnea Syndrome. Using 4 convolutional layers in their CNN, a ResNet-18, and XGBoost, they achieved an average specificity of 82\%.

In the 2017 ICBHI Challenge, \cite{icbhi} published a dataset of 920 lung sounds for finding the best method to identify abnormalities within the lung. Since the dataset's publication, there have been numerous attempts to identify the best classification method. \cite{mridha_respiratory_2021} set a new baseline for this dataset. While no data augmentation was performed, they chose to extract MFCCs to train their CNN consisting of 3 convolutional blocks. With a learning rate of .001, a momentum value of .8, a batch size of 125, and a dropout value of .03, they achieved a training accuracy of 98\% and a testing accuracy of 95\%. Following this work, in 2023, \cite{bapat_respiratory_2023} expanded on \citet{mridha_respiratory_2021} by adding new features to the model. While \citet{bapat_respiratory_2023} maintains a similar architecture to \citet{mridha_respiratory_2021}, the addition of Mel Spectrograms and Chroma Short Time Fourier Transforms (CSTFTs) improves the loss by 43\%.

\vspace{-\baselineskip}
\newpage
\section{Methods}

\hspace{1em} 

\begin{figure}[H]
   \centering 
   \includegraphics[height= 1.75in, width = 6in]{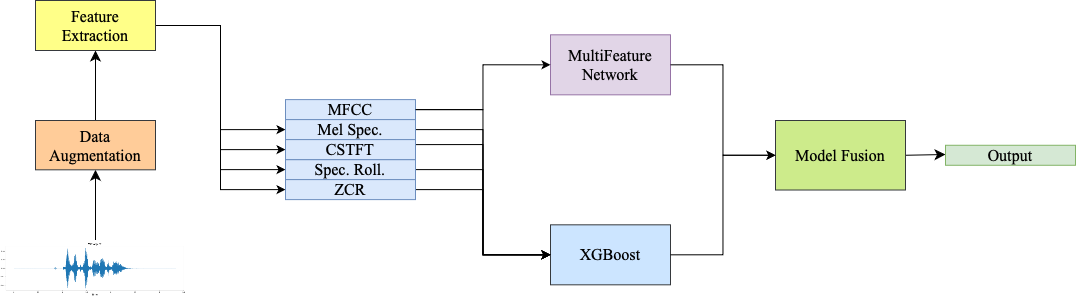} 
   \caption{We start by loading the audio file, then apply data augmentation and feature extraction. Next, we use the data to train the XGBoost Model and the Multifeature Network. Finally, we fuse the models together using soft voting to obtain the final classification.
}
   \label{fig:ens} 
\end{figure} 

\begin{figure}[!ht]
    \centering
    \includegraphics[trim={0 200 0 0},clip,width = 6in]{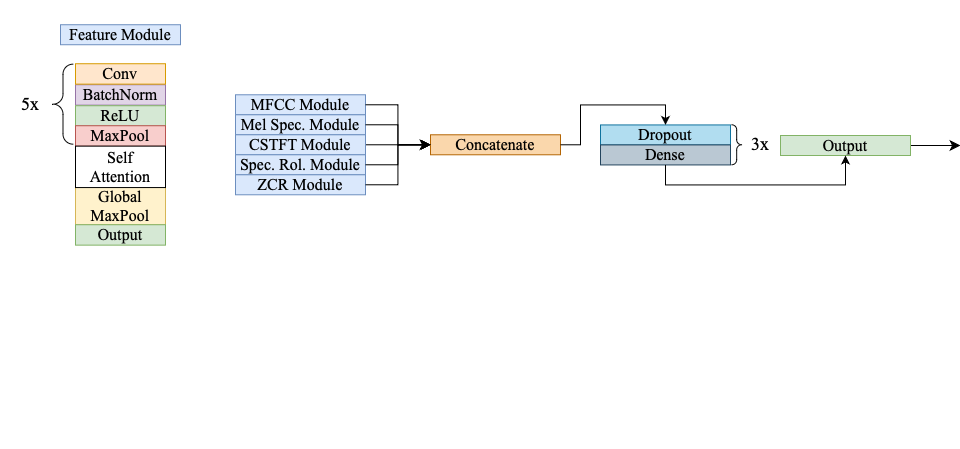}
    \caption{The Multifeature Network comprises of five individual CNNs for each respective feature. We introduce a new network consisting of 5 convolutional blocks, followed by a self-attention layer and a global max-pooling layer. We concatenate the outputs of each Feature Module and feed them to 3 dropout-dense blocks before outputting a prediction.
    }
    \label{fig:cnn}
\end{figure}
We introduce a novel ensemble learning approach for respiratory audio classification, leveraging a MultiFeature CNN and an XGBoost Model. In this section, we present our formulation of audio classification as an ensemble learning problem and discuss the rationale behind integrating multiple models. \ref{Architecture Overview} outlines the general pipeline of our approach, which combines the strengths of the MultiFeature CNN and the XGBoost Model. Then, we introduce an ensemble model for this task (\ref{Multifeature}). In Section \ref{XGB}, we introduce our XGBoost Model, which acts as a complementary classifier to the MultiFeature CNN.

\subsection{Architecture Overview} \label{Architecture Overview}

\hspace{1em} Building upon previous approaches to Audio Classification \cite{bapat_respiratory_2023, mridha_respiratory_2021}, a typical architecture consists of four components: Data Augmentation, Feature Extraction, Training, and Evaluation. Our goal is to maintain this foundation while introducing necessary complexity to enhance accuracy and efficiency. To expand our feature set, we incorporate the Zero Crossing Rate \cite{gouyon_use_2002} and Spectral Rolloff \cite{kos_acoustic_2013} into our feature extraction process. Additionally, we retain the importance of Data Augmentations \cite{alghamdi_deep_2024, chambres_automatic_2018} employing Bandpass Filters, Pitch Shifts, Shifts, and Additive White Gaussian noise.

Each feature is modeled separately within the Multi-feature CNN \cite{bapat_respiratory_2023}. Although their input tensors may vary, the core architecture of each CNN remains consistent. Subsequently, the outputs of the individual 2D CNNs \cite{mridha_respiratory_2021} are concatenated, and the model is compiled \cite{bapat_respiratory_2023}. Our CNN model incorporates a Self-Attention Mechanism to improve performance. This mechanism enables the network to focus on relevant features within the audio data by assigning varying degrees of importance to different input segments \cite{heitmann_deepbreathautomated_2023}.

We introduce the XGBoost model to enhance the classification performance of the Ensemble Network. The XGBoost constructs a series of decision trees sequentially, with each subsequent tree learning from the errors of the previous ones \cite{chen_xgboost_2016}. This iterative process allows the model to prioritize difficult-to-predict data points, gradually enhancing its predictive accuracy.

Following \cite{song_ahi_2023}, we employ soft voting to combine the predictions of the Multifeature CNN and XGBoost in a weighted manner. Unlike hard fusion, where the final prediction relies solely on the output of a single best-performing model, soft fusion assigns weights to each model's predictions and aggregates them to produce the final prediction \cite{akyol_ensemble_2024}.

\subsection{Multi-Feature Network} \label{Multifeature}

\hspace{1em} We introduce a novel sequential model for each feature CNN comprising five two-dimensional convolutional layers. For the MFCC model, input data is shaped as (40, 259, 1), where 40 denotes the number of MFCCs \cite{aykanat_classification_2017}, 259 represents the length of the temporal dimension, and 1 signifies the mono channel. Using this precedent, we apply the same rationale for deciding the input size for all additional features. Beginning with a sequence of five two-dimensional convolutional layers, we apply Batch Normalization, Rectified Linear Unit (ReLU) activation functions, and Max Pooling. We integrate batch normalization layers after each convolutional layer to stabilize and accelerate training. Following each convolutional layer, we incorporate Two-Dimensional Max Pooling layers to perform spatial downsampling of feature maps, enabling the model to recognize patterns irrespective of their location in the input. ReLU activation is chosen to mitigate the vanishing gradient problem.

We progressively increase the number of filters while reducing spatial dimensions in constructing the architecture. We begin with 32 filters using a 5x5 kernel size and strides of (2, 3) in the initial layer to facilitate significant downsampling. The second layer follows the same pattern, increasing the filter size to 64, while adopting a 3x3 kernel size with strides of (2, 2). This configuration maintains the downsampling while enhancing the model's ability to capture finer spatial details. We then progress to 96 filters with a 2x2 kernel size in the next layer, preserving the input's spatial dimensions for detailed feature extraction. In the last two layers, we integrate filters with a 2x2 kernel size, preserving more information at each step compared to previous architectures. This architectural design ensures the preservation of an optimal amount of information at each step while maintaining efficient downsampling, contributing to enhanced model performance and feature extraction capabilities.

The Self Attention layer, positioned just before the global max pooling layer, captures long-range dependencies and extracts relevant features from the input data. This approach improves performance while preserving and enhancing contextual information before aggregating the extracted features.

\subsection{Gradient Boosting with XGBoost} \label{XGB}

\hspace{1em} In this section, we introduce the XGBoost Model for respiratory disease classification. The XGBoost Model, an Ensemble Model employing decision trees and regularization techniques \cite{chen_xgboost_2016}, optimizes loss minimization through gradient descent during training. Unlike bagging techniques like Random Forest, boosting sequentially builds decision trees, with each subsequent tree aiming to reduce the error of the previous one. Choosing the number of trees is a delicate balance, as reducing the number of trees potentially leads to underfitting, whereas increasing the amount can lead to overfitting \cite{chen_xgboost_2016}.

We train the classifier on concatenated feature vectors extracted from the various features collected. The model is configured for multiclass classification by setting the objective to multiclass softmax. This setup calculates class probabilities and assigns instances to the class with the highest probability. Multiclass log loss serves as our evaluation metric, providing a comprehensive assessment by penalizing incorrect predictions proportional to the discrepancy between predicted and true class probabilities. In our case, setting the number of estimators to 400 balances accuracy and prevents overfitting.

\subsection{Ensemble Learning}

We employ soft voting to combine predictions from the CNN and the XGBoost using weighted averaging, where weights reflect the confidence or reliability of each learner's predictions. Unlike hard fusion methods, which directly combine base learners' outputs, soft fusion allows nuanced integration, assigning higher weights to confident or accurate predictions while reducing weights for those with higher uncertainty. This approach enables the ensemble model to make more informed predictions.

\section{Dataset}
The Respiratory Sound database is a collection of 920 lung sounds created by \cite{rocha2018} for the development of new techniques to differentiate unique respiratory sounds . These recordings span over 6898 respiration cycles and were annotated by experts in the field to denote the presence of crackles, wheezes, a combination of both, or their absence. They were recorded via stethoscope and were processed using an audio editing software. The recordings vary in length, and can span from between 10 seconds to 90 seconds. Each file contains information regarding the location in the body this audio was taken from. The annotation of chest locations from which the recordings were acquired provides spatial context and allows us to explore the relationship between specific respiratory sounds and their location within the chest cavity. This dataset was published for public use during the ICBHI challenge in 2017. Since its release, the dataset has been primarily used for respiratory disease classification, or for crackle and wheeze detection.

\subsection{Dataset Selection} 

\hspace{1em} This dataset represents patients of all ages and was selected for its variety and the surplus of information that can be extracted from each audio file and subsequent annotation. The audio files were collected from unique chest locations: the trachea, the left and right anterior chest, the posterior, and the lateral points on the body in both clinical and non-clinical environments. Within this dataset, the following diseases are represented with varying frequency: Lower Respiratory Tract Infections (LRTI), Upper Respiratory Tract Infections (URTI), COPD, Asthma, Pneumonia, and Bronchiectasis. Respiratory conditions encompass a wide spectrum of diseases and are each characterized by distinct patterns of respiratory sounds and clinical observations. This variety enables the identification of unique acoustic signatures associated with different diseases and facilitates more precise diagnostic algorithms. 

\subsection{Data Extraction and Augmentation} \label{augment}
\begin{figure}
    \centering
    \includegraphics[height = 3in, width=6in]{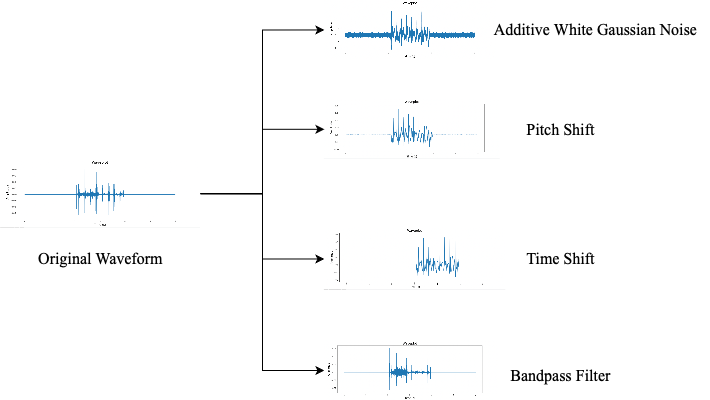}
    \caption{We apply AWGN, Bandpass Filters, Time Shifts, and Pitch Shifts to the waveform to increase the diversity of our dataset.}
\label{fig:aug}
\end{figure}
\hspace{1em} To address the inherent imbalance in the Respiratory Sound database, particularly with COPD cases outnumbering other respiratory conditions, we implemented a stratified splitting approach, as described in \cite{bapat_respiratory_2023}. This method ensured that each subset used for training, validation, and testing maintained a proportional representation of COPD cases alongside other diseases.

 We employed data augmentation techniques to enhance our model's capability and to mitigate potential biases. Augmenting the dataset with Additive White Gaussian Noise, Bandpass Filters, Shifts, and Pitch Shifts enabled us to simulate variations present in real-world respiratory sounds effectively. By introducing diversity into the dataset through augmentation, our models could learn from a broader spectrum of examples, reducing the risk of overfitting to specific instances. 

\begin{enumerate}
    \item Additive White Gaussian Noise (AWGN) 
    
    \hspace{1em} Additive White Gaussian Noise is a commonly used type of noise added to signals to simulate real-world interference. In our case, AWGN is particularly useful because it closely resembles background noise present in many recording environments. By incorporating AWGN during training, models become better equipped to handle noisy input.

    \item Bandpass Filter 
    
    \hspace{1em} Bandpass filters allow a specific range of frequencies to pass through while attenuating frequencies outside this range. This augmentation is valuable because bandpass filters maintain the original sample of the audio file, ensuring that the model encounters variations that occur in real-world scenarios.

    \item Time Shifts  
    
    \hspace{1em} Time shifting involves shifting the audio waveform along the time axis by a certain number of samples. In respiratory disease classification, shifts can be beneficial due to the variability in respiratory sounds across different individuals and conditions. Augmenting the training data with shifted versions of the original recordings helps the model become more robust to variations in onset times or duration of respiratory sounds.

    \item Pitch Shift

    \hspace{1em} Pitch shifting alters the pitch of the audio waveform while preserving its temporal characteristics. This augmentation is relevant because respiratory sounds may exhibit variations in pitch across different individuals, lung conditions, or recording conditions. By augmenting the training data with pitch-shifted versions of the original recordings, the model becomes more resilient to variations in pitch.

\end{enumerate}

\subsection{Feature Extraction}

\begin{figure} [H]
    \centering
    \includegraphics[height = 4.5in, width= 5in]{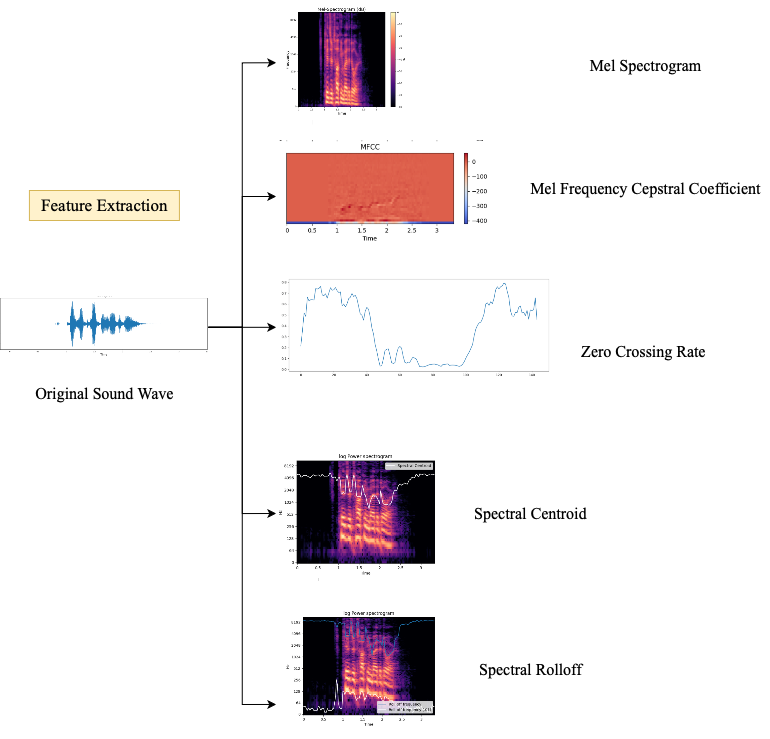}
    \caption{From the augmented data, we utilize the librosa library to extract MFCCs, CSTFTs, ZCRs, Mel Spectrograms, and the Spectral Rolloffs.}
\label{fig:extract}
\end{figure}

\hspace{1em} Feature extraction plays a crucial role as they capture unique aspects of the sound, encompassing important information such as frequency patterns, spectral characteristics, and temporal features inherent in each respiratory signal. In our approach, we selected Mel Frequency Cepstral Coefficients (MFCCs), Mel Spectrograms, Chroma Short Time Fourier Transforms (CSTFTs), following the methodology outlined in \cite{bapat_respiratory_2023}. Additionally, we expanded the feature set to include Spectral Rolloff \cite{kos_acoustic_2013} and the Zero Crossing Rate \cite{gouyon_use_2002}. We discovered that collectively, these features effectively highlight the unique abnormalities associated with each respiratory disease, such as wheezing or crackles, more so than their counterparts.

\begin{enumerate}
    \item Mel Frequency Cepstral Coefficients (MFCCs) \label{mfcc}
    
    \hspace{1em} MFCCs capture the spectral characteristics of audio signals by representing their short-term power on the Mel Scale, a perceptual pitch scale that mirrors human auditory response to frequency. These coefficients are instrumental in highlighting potential abnormalities by characterizing frequency distributions and their intensities.
    
    \item Chroma Short Time Fourier Transforms (CSTFTs) 

    \hspace{1em} Chroma Short Time Fourier Transforms (CSTFTs) illustrate the frequency content of audio signals over time, emphasizing specific frequency components. By capturing spectral variations over time, CSTFTs reveal the significance of transient events in the signal, allowing monitoring of the progression of adventitious sounds throughout the audio file.

    \item Mel Spectrograms 
    
    \hspace{1em}
    Similar to MFCCs, Mel Spectrograms offer a detailed depiction of the input's frequency content. By applying the Mel Scale to the waveform, we highlight important frequency bands relevant to the signal. Additionally, Mel Spectrograms provide a more compact representation of sound frequency, reducing dimensionality and feature space. This simplification streamlines the model training process and increases its efficiency. We choose to follow \cite{aykanat_classification_2017}'s optimal value for MFCCs per window at 40.

    \item Spectral Rolloff \label{roll}

    \hspace{1em} The Spectral Rolloff measures the frequency below which a certain percentage of total spectral energy occurs, dividing the spectrum into low and high-frequency regions. This feature emphasizes tonal differences across audio signals and can indicate unique harmonics associated with each disease.

    \item Zero Crossing Rate \label{zcr}

    \hspace{1em} The Zero Crossing Rate measures how quickly the audio signal alternates between positive and negative values within short time intervals. A higher ZCR suggests a more percussive sound, while a lower ZCR corresponds to smoother, sustained sounds. We employ ZCR to identify abrupt changes or irregularities in each signal.
    
\end{enumerate}

\section{Results} 

This section discusses an explanation of the architecture training model and shows our results. We also showcase the technology utilized in this research. Secondly, we display tables consisting of our results. 

In evaluating our model's performance, we draw inspiration from established metrics utilized in related works such as \cite{mridha_respiratory_2021}, \cite{bapat_respiratory_2023}, and \cite{chambres_automatic_2018}. These metrics have proven effective in assessing the efficacy of respiratory sound analysis systems and serve as benchmarks for our evaluation.

Initially, we employ fundamental metrics, including training and testing loss, and accuracy. We discuss our loss function further in \ref{res}. These metrics offer a basic understanding of the model's performance in terms of its ability to minimize errors and classify instances correctly. However, given the nature of the respiratory sound classification problem, we recognize the limitations of relying solely on accuracy. Therefore, we augment our evaluation with more comprehensive metrics. Precision, Recall, and AUC (Area Under the Curve) are essential metrics in classification tasks, and we choose them for this work. We choose these metrics as they offer a more holistic understanding of the model's performance compared to accuracy alone, making them valuable indicators for assessing the effectiveness of AFEN.
\subsection{Experimental Evaluation} \label{res}

\begin{table}[h]
\centering 
\begin{tabular} {|c|c|c|c|c|c|}
\hline
Model & Epochs & Training & Testing & AUC & Testing Loss \\
\hline 
MultiFeature CNN & \textbf{100} & .9959 & \textbf{.9827} & \textbf{.9963} & \textbf{.0869} \\
\hline 
XGBoost Model & 400 & .9831 & \textbf{.9726} & \textbf{.9847} & \textbf{.0367} \\
\hline 
Ensemble Model & 100 & .9971 & \textbf{.9757} & \textbf{.9943} & \textbf{.0532} \\ 
\hline 
\end{tabular}
\caption{Training, Testing, and AUC metrics for each model. XGBoost Runs on 400 estimators individually, but the ensemble trains in 100 epochs. The values in bold represent the state-of-the-art values achieved for each individual model.}
\label{t_combined}
\end{table}

From \ref{t_combined}, we set the MultiFeature CNN to 100 epochs and we remain consistent with \cite{bapat_respiratory_2023}, as we set the train-test-split to 80-20. Similarly, we choose to define sparse categorical crossentropy as our loss function for multiclass classification. 

\newcommand{\mysubscript}[1]{\raisebox{-0.34ex}{\scriptsize#1}}
After initially setting the epochs to 250 \cite{mridha_respiratory_2021}, we found the model converged to the state-of-the-art value for MultiFeature CNNs on this dataset at 98.27\% at 100 epochs. Additionally, we set the new state-of-the-art for testing loss at 8.69\%. For the XGBoost,we set the state-of-the art in testing accuracy at 97.26\% and a testing loss of 3.67\%. In setting the number of estimators to 100, we received a test accuracy of 93\%, far below our expectations compared to the MultiFeature CNN. Due to the size of our dataset, we had to increase the number of estimators. We choose 400 estimators only after meticulously, and iteratively, increasing and decreasing the number of estimators to achieve our expected accuracy while preventing overfitting. Additionally, we maintain the default learning rate at .3 and choose multiclass logloss as our loss function (\ref{XGB}). This is defined as:

\begin{equation}
L\mysubscript{log} (Y,P) = -logPr(Y|P) = -\frac{1}{N} \sum_{i=0}^{N-1} \sum_{k=0}^{K-1} y\mysubscript{i,k}log(p\mysubscript{i,k})
\end{equation}

Where 
\begin{itemize}
    \item \( Y \): True labels or ground truth values, encoded in a one-hot format.
    \item \( P \): Predicted probabilities outputted by the model for each class.
    \item \( \text{Pr}(Y|P) \): Probability of observing the true labels given the predicted probabilities.
    \item \( N \): Number of samples or data points in the dataset.
    \item \( K \): Number of classes in the classification problem.
    \item \( y_{i,k} \): Element in the true label matrix \( Y \) corresponding to the \( i \)-th sample and the \( k \)-th class.
    \item \( p_{i,k} \): Element in the predicted probability matrix \( P \) corresponding to the \( i \)-th sample and the \( k \)-th class.

\end{itemize}

\begin{table}[h]
\centering
\small
\setlength\tabcolsep{6pt}
\begin{tabular}{|l|c|c|c|c|c|c|}
\hline
\rotatebox{90}{Class} & \rotatebox{90}{Precision (CNN)} & \rotatebox{90}{Recall (CNN)} & \rotatebox{90}{Precision (XGB.)} & \rotatebox{90}{Recall (XGBoost)} & \rotatebox{90}{Precision (Ensemble)} & \rotatebox{90}{Recall (Ensemble)} \\
\hline 
Asthma & 1.00 & .17 & 1.00 & .33 & 0.00 & 0.00 \\
\hline 
Bronchiectasis & .94 & .94 & .98 & .80 & 1.00 & .94 \\
\hline 
Bronchiolitis & .90 & .89 & .95 & .76 & .96 & .93 \\ 
\hline 
COPD & .99 & .99 & .98 & 1.00 & 1.00 & 1.00 \\
\hline 
Healthy & .79 & .93 & .90 & .89 & .94 & .96 \\ 
\hline 
LRTI & .86 & .75 & .96 & .72 & .88 & .76 \\ 
\hline 
Pneumonia & .99 & .79 & .94 & .89 & .98 & .91 \\ 
\hline 
URTI & .83 & .85 & .90 & .85 & .90 & .86 \\
\hline 
\end{tabular}
\caption{Comparison of Precision and Recall for Multifeature CNN, XGBoost, and Ensemble Models. }
\label{disease_results}
\end{table}

The Multifeature CNN demonstrates exceptional precision across various classes, most notably in COPD and Pneumonia. The classes with higher performance, such as COPD, Pneumonia, and Bronchiectasis, correspond to higher support levels, while those with lower support display greater volatility in the model's predictions. Similarly, the XGBoost model showcases high precision scores, exceeding 0.90 for classes like Asthma, Bronchiectasis, and COPD. It builds on the Multifeature CNN in that it demonstrates strong recall scores, especially for COPD, achieving perfect recall. Overall, both models maintain robust precision and recall metrics across various classes.

Comparing the precision and recall table for the Ensemble Model with those of the previous models reveals notable improvements in performance across various classes. The Ensemble Model consistently achieves high precision scores across all classes, with classes. Moreover, it excels in recall scores across the majority of classes, particularly in categories such as Bronchiectasis, COPD, and Healthy. While individual models exhibited high precision scores, the Ensemble Model notably enhances recall values across most classes, with improvements ranging from minor to substantial, barring Asthma, which lacks sufficient support and would benefit from additional unique data to bolster its scores.

\section{Discussion} 

 \hspace{1em} This study introduces AFEN, an ensemble learning methodology for respiratory disease diagnosis. To achieve this, we introduce a novel data processing pipeline that integrates a unique data augmentation process and an expanded feature set. We advance the capabilities of Multifeature CNNs by introducing self-attention mechanisms within the feature CNNs. By leveraging self-attention mechanisms as in \cite{heitmann_deepbreathautomated_2023} and fusing the output of the Multifeature CNN with the XGBoost gradient boosting framework, we observe a significant improvement in performance compared to \cite{chambres_automatic_2018}, \cite{mridha_respiratory_2021}, \cite{bapat_respiratory_2023} and lead to more accurate and reliable diagnoses.

 Our work represents a breakthrough in machine learning-driven audio classification, extending beyond healthcare boundaries. The methodologies developed in this study can be adapted and extended to various audio classification tasks. The high precision and recall scores achieved by the Ensemble Model indicate its potential as a reliable tool for aiding medical professionals in diagnosing respiratory conditions, leading to more accurate and timely diagnoses.

The model not only demonstrates exceptional accuracy in classification, but also proves to be a valuable real-time diagnostic tool. Its high classification scores highlight its potential to diminish the need for invasive diagnostic procedures, saving time and resources. By offering rapid and accurate assessments, the model equips healthcare practitioners with timely insights for decision-making and treatment planning. Moreover, its non-invasive nature compared to medical imaging makes it a less harmful diagnostic medical device, while also being more accessible and cost-effective. 

\paragraph{Limitations}

While our approach demonstrates promising results, there are limitations that need to be acknowledged. Firstly, the generalizability of our findings may be limited by the specific characteristics of the dataset used in this study, which could affect the performance of the models when applied to new and unseen data. Moreover, ensemble learning involves multiple models which can result in increased computational requirements and a larger memory footprint of the generated model. Additionally, the reliance on medical data obtained from a single institution or source may introduce biases or limitations in the model's ability to generalize to different populations or healthcare contexts. When working with imbalanced datasets, particularly in clinical contexts, the recall metric becomes particularly sensitive to data distribution variations and the occurrence of false negatives. In scenarios where certain diseases, such as Asthma in our case, are less prevalent or inadequately represented in the training data, the model's ability to generalize and accurately predict them is challenged. As a result, the recall scores for these classes exhibit notable volatility. 

\bibliography{audioref}

\newpage
\section*{Appendix}
    \begin{figure}[H]
      \centering
      \includegraphics[height = 2in, width = 4in]{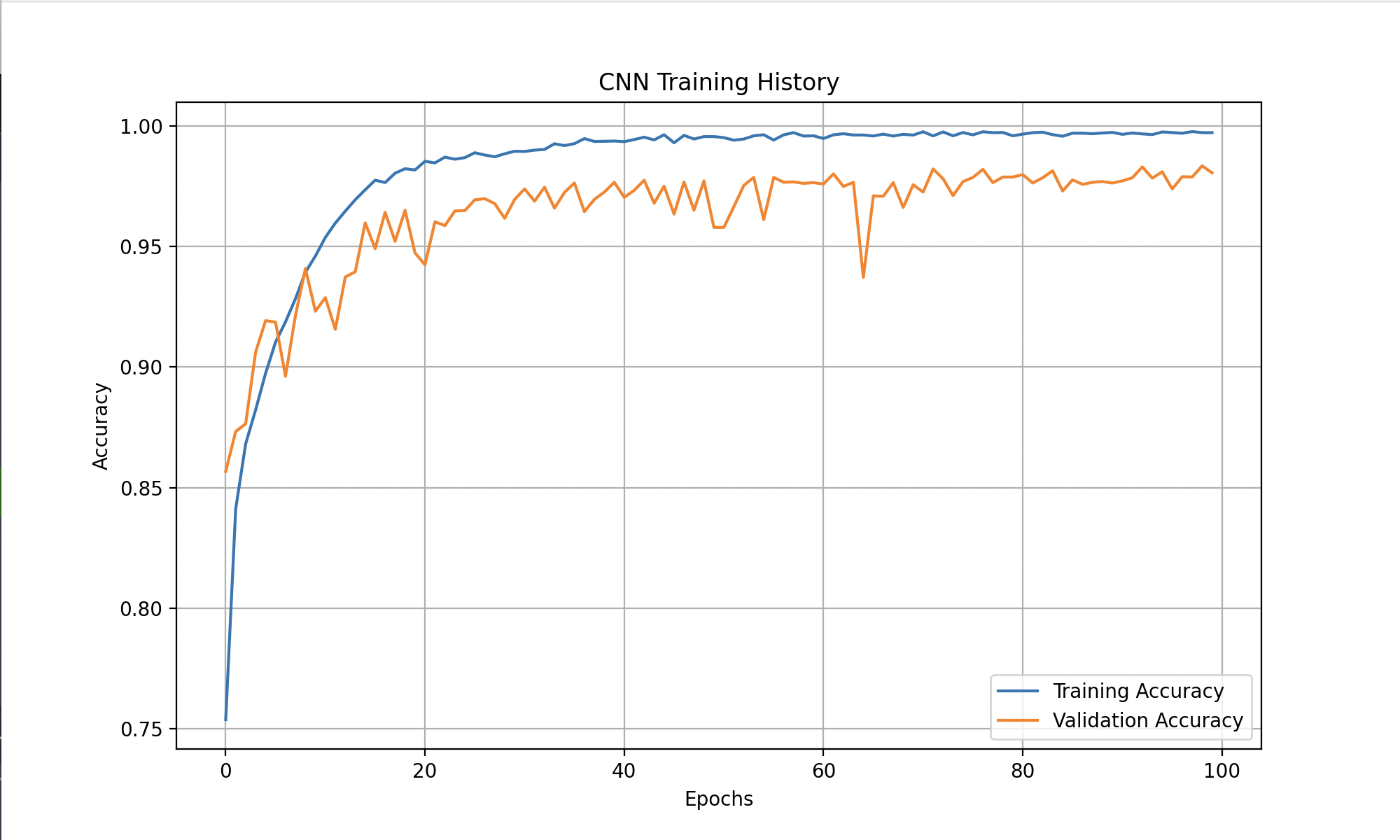}
      \caption{Training and Validation Accuracy for MultiFeature CNN}
      \label{fig:sub1}
    \end{figure}
    
    \begin{figure}[H]
      \centering
      \includegraphics[height = 2in, width = 4in]{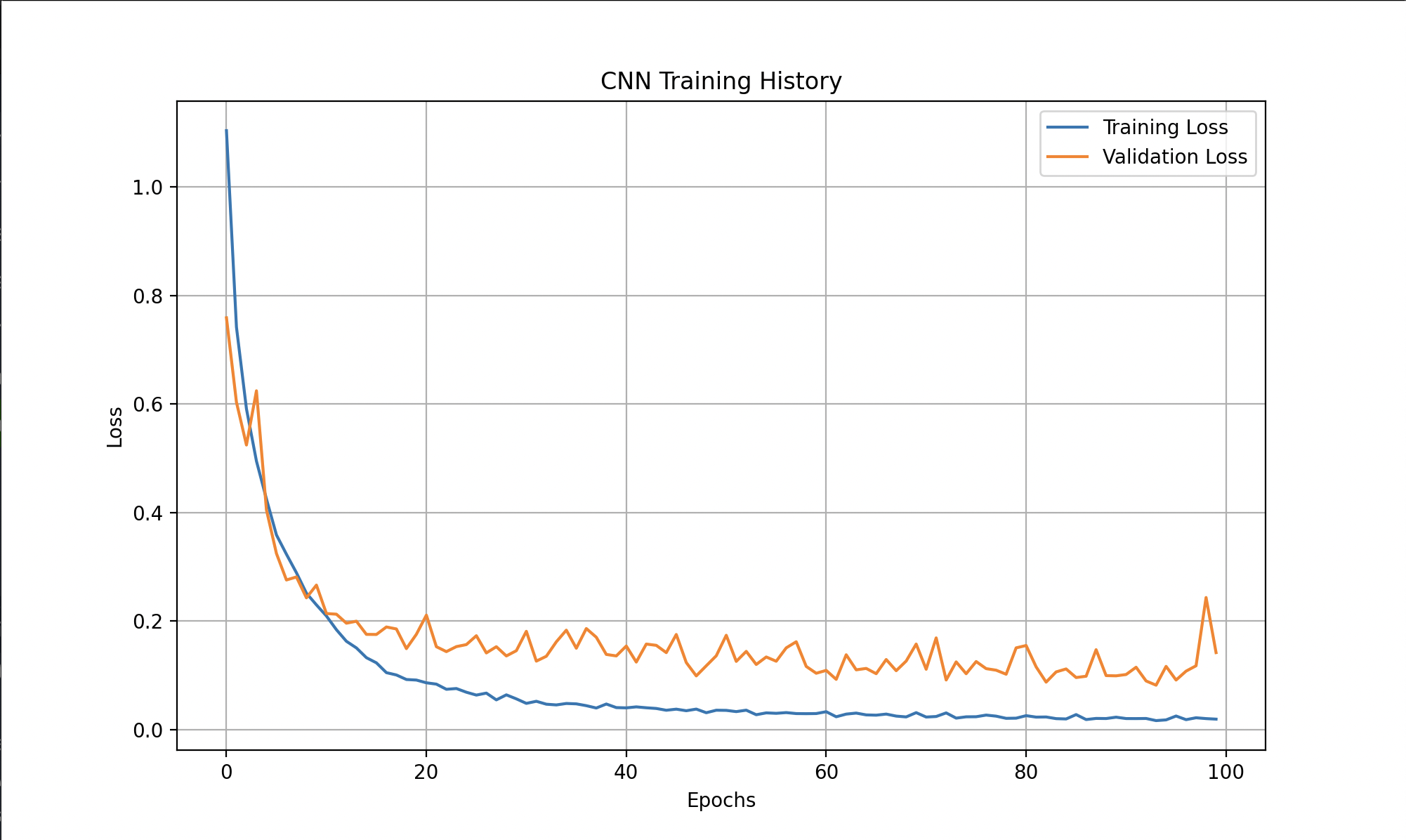}
      \caption{Training and Validation Loss for MultiFeature CNN}
      \label{fig:sub2}
    \end{figure}
    
    \begin{figure}[H]
      \centering
      \includegraphics[height = 2in, width = 4in]{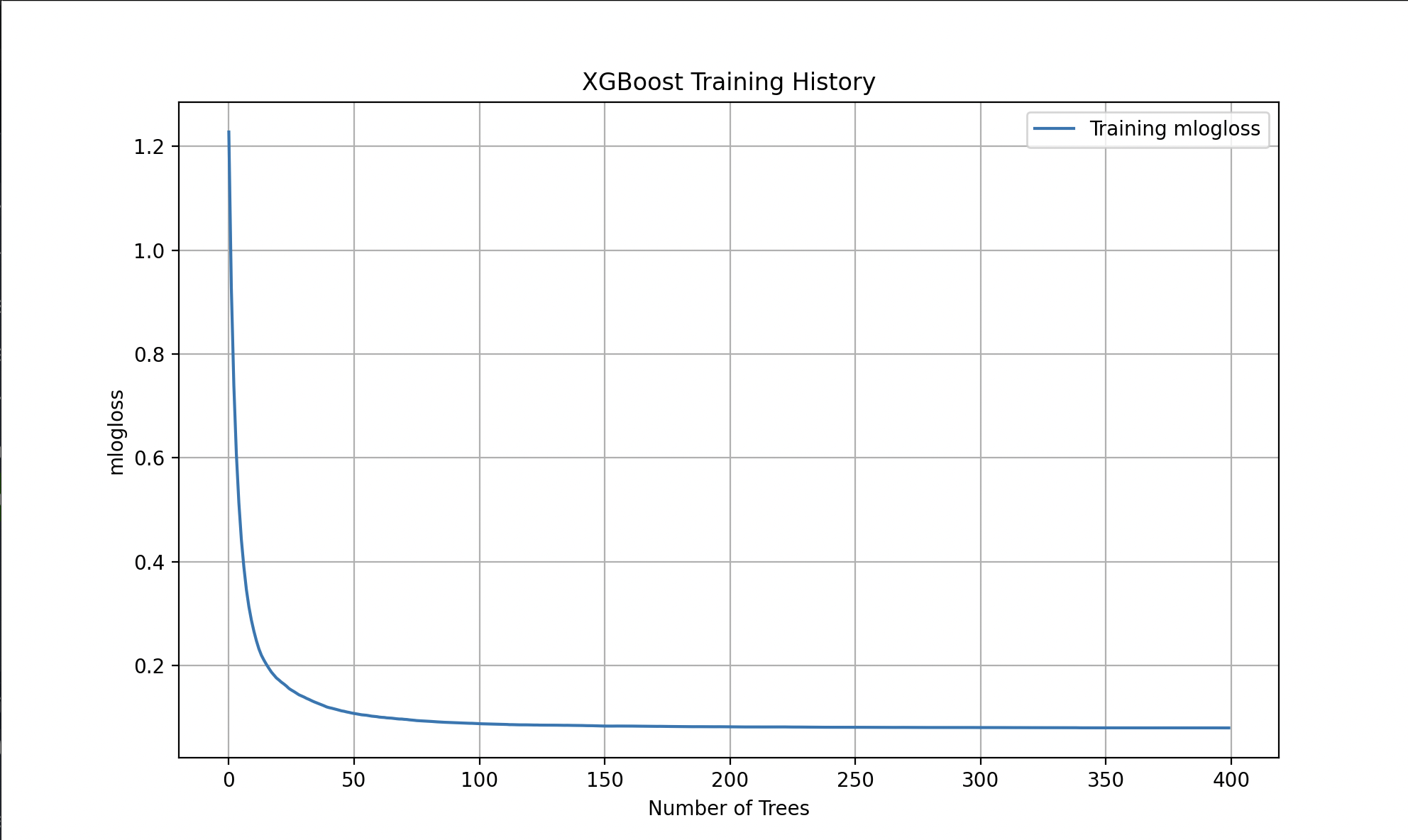}
      \caption{Training loss for the XGBoost Model}
      \label{fig:sub3}
    \end{figure}
    
\end{document}